\documentclass[conference]{IEEEtran} 

\usepackage{comment}
 \newtheorem{definition}{Definition}
\usepackage{booktabs}

 \usepackage{tcolorbox}
    \usepackage{algpseudocode}
\algnewcommand\algorithmicforeach{\textbf{for each}}
\let\oldReturn\Return
\renewcommand{\Return}{\State\oldReturn}
\algdef{S}[FOR]{ForEach}[1]{\algorithmicforeach\ #1\ \algorithmicdo}

\newcommand{\ihead}[1]{\par\noindent\textit{#1} --}

\usepackage{amsmath,epsfig,epstopdf} 

\newcommand{\head}[1]{\par\noindent\textbf{#1:}\space}

\newcommand{\headQ}[1]{\par\noindent\textbf{#1}}

\usepackage{lipsum}

\IEEEoverridecommandlockouts 

\makeatletter
\def\thanks#1{\protected@xdef\@thanks{\@thanks
        \protect\footnotetext{*#1}}}
\makeatother

\makeatletter
\renewcommand\footnoterule{\relax\kern-5pt
\hrule
\kern4.6pt}
\makeatother

\usepackage{float}
\usepackage{glossaries}
\floatstyle{boxed}
\newfloat{algorithm}{htbp}{loa}
\floatname{algorithm}{Algorithm}
\newfloat{algorithm}{htbp}{loa}

\usepackage{multicol,tabularx}
\usepackage{array}
\usepackage{makecell}

\usepackage{siunitx}

\usepackage[utf8]{inputenc}
\usepackage{booktabs}
\usepackage{tcolorbox}
\usepackage{amsmath}
\usepackage{nccmath}
\usepackage{tabularx}
\usepackage{amssymb}
 \usepackage{graphicx}
\usepackage{amsmath}
\usepackage{multicol,tabularx}
\usepackage{color, colortbl}
\definecolor{lightgray}{gray}{0.9}
\definecolor{Gray}{gray}{0.9}
\usepackage{array}
\usepackage{makecell}
\usepackage{multirow}
\usepackage{csvsimple}

\title{On The Effectiveness of Program Reduction Techniques in Automated Program Repair}

\author{Omar I. Al-Bataineh \\ Jordan University of Science and Technology}

\date{}

\begin{document}

\maketitle

\begin{abstract}

 Repairing a large-scale buggy program using current automated program repair (APR) approaches can be a difficult time-consuming operation that requires significant computational resources.
We describe a program repair framework that effectively handles large-scale buggy programs of industrial complexity. The framework exploits program reduction in the form of program slicing  
to eliminate parts of the code irrelevant to the bug being repaired without adversely affecting the capability of the repair system in producing correct patches.

Observation-based slicing is a recently introduced, language-independent slicing technique that shows a good effectiveness in a wide range of applications.
In this work, we show how ORBS can be effectively integrated with APR to improve  all aspects of the repair process including the fault localisation step,
 patch generation step, and  patch validation step. 
 The presented repair framework indeed enhances the capability of APR by reducing the execution cost of a test suite
 and the search cost for the appropriate faulty statement corresponding to the bug being repair. 
 Our empirical results on the widely used Defects4J dataset
reveal that a substantial improvement in performance
can be obtained without any degradation in repair quality.
\end{abstract}

\section{Introduction}

Automated program repair (APR) is an active research area with expanding tool support~\cite{monperrus8,GouesPR19}.
However, 
APR has struggled with scalability and grows time-consuming and resource-intensive for larger programs.  
This paper aims to enhance APR performance by exploiting program reduction in the form of program slicing to reduce both the search time for the faulty statement and  validation times for candidate patches. 

APR consists of three steps: fault localization, patch generation, and patch validation.
Poor performance of the first phase, fault localization (FL) can lead to poor overall performance when the actual faulty statement is found far down the list of suspicious statements~\cite{Liu20,Liu19FL}. 
Poor ranking impacts APR performance by increasing the number of failed patch attempts.
Another aspect that contributes to the poor performance of APR tools is the cost of patch validation, which is heavily reliant on the test suite size.
Experience has shown that much of a test suite is irrelevant to fixing a given bug.
The expense of running these tests decrease APR's efficiency.
This paper experiments with test suite reduction and suspicious list optimisation
that make use of the program slice.

To demonstrate the effectiveness of combining slicing with APR,
we apply TBar~\cite{LiuK0B19}, a state-of-the-art template-based APR tool, to selected bugs form the widely used Defects4J dataset~\cite{JustJE14}. 
We use template-based APR because it is one of the most effective approaches to generating patches~\cite{Liu20}.
We use the bugs from Defects4J to illustrate the effectiveness of our approach and to consider its scalability.
The empirical results reveal that a substantial improvement in performance can be obtained without any degradation in repair quality.

\head{The need for accelerating APR}
Real-world programs grow ever larger.   For example, the modern vehicles include millions of lines of code.
In spite of the successes of APR in fixing a wide variety of bugs in real-world programs, additional research is needed to handle programs of industrial complexity. 
Current APR approaches simply take too long.
For example, in the benchmarks we study even a single line repair can take hours and in several cases the search fails because of the patch space is too large.
While we study template-based repair, recent NN-based approaches are even slower \cite{ZhangFMSC24}. For example, the recent NN-based tool recoder consumes more than 4 hours on average to fix the bugs of Defects4J.
The main reason for this slow running time is 
APR tools commonly must consider a large number of patch candidates before finding a valid patch. 
To expedite the APR process, acceleration approaches based on the concepts of program reduction techniques are needed.

\head{Demonstrating example} As a preliminary illustration, consider the repair of Defects4J~\cite{JustJE14} bug Lang-10 using the state-of-the-art template-based APR tool TBar~\cite{LiuK0B19}.
TBar takes seven hours to generate a plausible patch.
The long processing time can be explained by the following observations.
First, TBar spends a significant time mutating suspicious statements that are not the actual 
faulty statement responsible for the bug:  the fault localization results for Lang-10 are a ranked list of 287 suspicious statements, in which the faulty has rank 70. 
Second, the test suite for bug Lang-10 consists of 2196 passing tests and 2 failing tests where most of the tests are irrelevant to Lang-10 
and thus can be skipped without adversely affecting the quality of generated patches.
A minimised test suite obtained by removing irrelevant tests using a slicing-based test reduction approach consists of only 52 tests.
Using the minimised test suite we discover the same patch is produced after examining 1476 patches in approximately 40 minutes.

\head{Contribution}
 We show how slicing can be \textit{effectively} integrated with APR to improve all aspects of the repair process
    including the FL step, the patch generation step (the number of examined patch candidates), and the patch validation step. 
    We show first how ORBS slicer \cite{binkley2014} can be used to optimise FL by improving the suspiciousness rank of the actual faulty statement corresponding to the target bug,
    which in turn reduces the number of examined patch candidates.
   We also show how slicing can be used to reduce the size of test suites  by eliminating irrelevant tests without adversely affecting the quality of produced patches. 
 We demonstrate the effectiveness of our repair framework approach that combines the ORBS program slicing with APR across the widely used Defects4J benchmark~\cite{JustJE14}, demonstrating promising results using ORBS to improve fault localization effectiveness, as well as the patch generation and patch validation steps of APR. 

\head{Notations} Throughout the paper we use the following notations. 
We write $P$ to refer to the original program containing bug $b$,
$P_S$ to refer to the corresponding sliced version of $P$, $T$ a test suite for $P$, $T_R \subseteq T$ a reduced test suite obtained by eliminating irrelevant tests of $T$ w.r.t. bug $b$, 
$SL$ the list of suspicious statements obtained by running an FL technique using $P$ and $T$, and $SL_R$ is a reduced  suspicious list obtained for bug $b$ using $P$ and $T_R$. 
Finally, we refer to $R (P, T, SL)$ as a \textit{standard repair setup} in which the original program $P$, the entire test suite $T$, and the list $SL$ are used when running the repair system $R$ to produce a patch for bug $b$.

\section{Background}

\noindent
This section introduces APR's efficiency metrics that we use in this work and observation-based slicing (ORBS).

\head{Efficiency metrics for APR}
We use two efficiency metrics to measure the efficiency gain brought on by the use of slicing: (i) the repair time (RT) or the time it takes the APR tool to generate a patch,  and (ii) the number of patches examined before a valid patch is found (known as the NPC score \cite{QiMLW13}). 
While reducing the size of the test suite is expected to improve the RT but not the NPC, improving the suspiciousness rank of the faulty statement is expected to improve both RT and NPC.
The effectiveness impact of slicing on the performance of APR tools needs therefore to be assessed using both metrics. 

\head{Observation-based slicing} 
The following description of Observation-Based Slicing (ORBS) is taken from
existing work on the technique~\cite{binkley2014, islam:porbs16, stievenart2023empirical}.

The key to ORBS is \emph{observation}.
ORBS takes as input a source program $P$, a slicing criterion identified by a program variable $\nu$, a program location $l$, a set of inputs $\mathcal{I}$, and a maximum window size $\delta$.
The resulting slice preserves the values of $\nu$ at $l$ for the inputs of $\mathcal{I}$.

ORBS \emph{observes} behavior by instrumenting $P$ with a side-effect free line that tracks the value of variable $\nu$ immediately before line $l$.
Then, starting with $P$, ORBS repeatedly forms candidate slices by deleting a sequence of one to $\delta$ lines from the current slice.
The candidate is rejected if it fails to compile or produces different values for $\nu$.
Otherwise, the candidate becomes the current slice.
ORBS systematically forms candidates until no more lines can be deleted.


\section{APR and Reduction Techniques}

The program reduction strategy we study in this paper is dynamic program slicing.
A dynamic slice of program $P$ is a syntactic subset of $P$ that preserves the behavior of a given part of the code called a \emph{slicing criteria} on a select set of $P$'s tests.
Slicing's goal is to produce the simplest such program.
Applied in the context of APR, slicing is useful as it abstracts away code irrelevant to the bug being repaired and thus reduces the effort required by the APR tool. 
We next formalize the requirements that a slice must satisfy to be useful in APR.
%
\begin{definition} \textbf{Feasible slices}\label{FeasibleSlices}.
Let $P$ be a program that contains a statement $s$ which is the location of bug $b$ 
and let $P'$ be a slice of $P$ taken with respect to $s$ and the variables used in $b$ (i.e., $s$ and the variables used in $b$ form the slicing criterion).
Program $P'$ is a \emph{repair-feasible} slice of $P$ iff
\begin{enumerate}
    \item  $P'$ is a smaller executable version of $P$, and

    \item   all runs of $P$ that exhibit bug $b$ have a corresponding run of $P'$ that also exhibits $b$.
    
    
\end{enumerate}
\end{definition}

The first property of Definition~\ref{FeasibleSlices} concerns the \textit{efficiency} of the APR process: it is essential that repairing the sliced program will be faster than repairing the original program.
The second property concerns the \textit{accuracy} of the applied slicing method: if the slicing method skips some of the erroneous behaviors in the original program then the resultant slice is no longer feasible for repairing the target bug.

\subsection{Reducing Test Suites by Eliminating Irrelevant Tests}

Most state-of-the-art APR tools use test suites to localize the faulty statement of the bug under repair, and then generate and validate the resulting candidate patches. 
Tests used by APR typically consists of two parts: (i) failing tests $T_F$ used to trigger the defective behavior in the program, 
and (ii) passing tests $T_P$ used to model the expected correct behavior of the program. 
The number of failing and passing tests employed in the APR process affects not only the quality of the generated patches but also the accuracy  of the fault localization step.

However, test suites written for large-scale programs  
are often constructed without developing sufficient knowledge about the dependency relationships between the different components of the program under repair.
Thus, a 
test suite may contain tests that are irrelevant to the bug being repaired.
The repeated execution of these irrelevant tests during patch validation can significantly degrade the performance of APR tools.
There is, therefore, a need to identify and eliminate irrelevant test cases in an test suite (see Definition \ref{Irrelevant tests}).

\begin{definition} \textbf{Identifying irrelevant tests}\label{Irrelevant tests}.
Let $P$ be a program containing a bug $b$ and
$S_f$ be the set of faulty statements responsible for the occurrence of $b$ and $T$ be a test suite for program $P$.
We say that a test $t \in T$ is an irrelevant test w.r.t. bug $b$ if $t$ tests behaviors in $P$ (i.e. a sequence of statements $S$) that are semantically independent from $S_f$. That is, the computations of $S_f$ and $S$ are mutually irrelevant.

\end{definition}

\subsection{Optimising Suspicious Lists by Reducing False-Positives}

Despite significant advances~\cite{Abreu07,Zhang07,Zhang05,Zhang11,Xuan14,LidO16,Pearson17,LouGLZZHZ20,Li0N21a}, FL still suffers from accuracy issues~\cite{Liu20,Liu19FL}.
For example, the top ranked suspicious statements are often false positives. 
This poor ranking of faulty statements directly affects APR performance by
increasing the number of patch candidates that are needlessly considered. 
Therefore, valid optimisation and reduction techniques must be applied to reduce the number of false-positives in the suspicious lists produced by current FL techniques (see Definition \ref{ValidSLReduction}).

\begin{definition} \textbf{Valid suspicious list optimisation}\label{ValidSLReduction}.
Let $P$ be a program containing bug $b$,  $S_f$ be the set of faulty statements responsible for the occurrence of $b$, and $T= (T_F \cup T_P)$ be a test suite developed for testing $P$. 
Let also $P_S = ORBS (P, T_F)$ be a program slice of $P$ computed by running the set of failing tests $T_F$, and $T_R \subseteq T$ be a reduced test suite obtained as described at Definition \ref{Irrelevant tests}.
For an FL  algorithm $F$, we say that $ SL_R = F (P, T_R)$ is a \emph{valid optimised} suspicious list of $P$ iff:
 
 \begin{enumerate}
     \item the reduced suspicious list  $SL_R$ contains $S_F$, and
     \item the setup $ R (P, T, SL_R)$ produces a valid patch for $b$ or one that is identical to that created by $R(P, T, SL)$, and
     \item  NPC score of  $R (P, T, SL_R)$ is smaller than that of $ R (P, T, SL)$. That is, $R$ examines a fewer number of patches before finding a valid one using  the list $SL_R$.
 \end{enumerate}
\end{definition}

In Subsection \ref{sec:empiricalResults}, we empirically evaluate the effectiveness of ORBS in identifying and eliminating irrelevant tests, as well as lowering false-positives in generated suspicious lists.

\section{Optimised Template-based Repair Algorithm}

Our repair algorithm (see Algorithm \ref{alg:OptimizedAlgor}) aims to improve all aspects of APR
including the fault location step, the patch generation step, and the patch validation step. 
We do this using slicing, which typically reduces the size of the test suite and suspicious list as well. 
  Statements that are not part of the slice can be removed from the 
  original suspicious list thus reducing the number of false positives.
This in turn helps reduce the number of irrelevant patches considered. 
Additionally, slicing can help identify and eliminate irrelevant tests from the test
suite, which reduces the cost of  the patch
validation step. 

More formally, our algorithm consists of the following phases applied to buggy program $P$ and its test suite $T$. 

\begin{enumerate}

    \item \textit{Program reduction phase}. This phase produces a slice $P_S$ of $P$ by eliminating code irrelevant to the bug.
    
    \item \textit{Test suite reduction phase}. This phase constructs a reduced test suite $T_R$ by removing all tests from $T$
          that test parts of $P$ not found in $P_S$.
    
    \item \textit{Suspicious list optimisation phase}. This phase runs GZoltar \cite{Campos12} on $P$ using $T_R$ to produce
         the list $SL_R$.
    
    \item \textit{Repair phase}.  This phase applies $R$ to $P$, $T_R$, and $SL_R$.
\end{enumerate}

\begin{algorithm} 
\small
\begin{algorithmic}[1]
\State \textbf{Inputs}: The project to be repaired $P$ and its test cases $T$
\State \textbf{Output}: the patch candidate $pt$ that passes all test cases
\State $pt:= NULL$
\State $P_S := ORBS (P, T) $ 
\State $T_R := TestMinimisation (P_S, T)$
\State $SL_R := fault\_localisation (P, T_R)$
\ForEach {$loc \in SL_R $}
\ForEach{$ft \in fix\_templates$}
\State $donor\_code\_list := search\_donor\_code (loc, ft)$
\ForEach{$item \in donor\_code\_list$}
\State $candidate := patch\_generate (loc, ft, item)$
\If{$validate (candidate, T_R)$}
\State $pt := candidate$
\Return $pt$
\EndIf
\EndFor
\EndFor
\EndFor
\Return $pt$
\end{algorithmic}
\caption{Optimised template-based repair process}
\label{alg:OptimizedAlgor}
\end{algorithm}

The effectiveness of test suite and suspicious list reduction is relative 
to the soundness of the generated  slice corresponding to the original buggy program. 
The number of deleted irrelevant statements and deleted irrelevant tests depends directly on the effectiveness of the employed  reduction technique.

\section{Research Questions}


During the analysis we consider the following questions:

\ihead{RQ1} How effective is ORBS at reducing test suites?

\ihead{RQ2} How effective is ORBS in improving FL results?

\ihead{RQ3} Does improving FL improve the performance of TBar?

\ihead{RQ4} Does reducing test suites  improve TBar's efficiency?


\section{Evaluation}

The evaluation makes use of a spectrum-based fault localization (SBFL) technique in the latest version (v1.7.2) of GZoltar, the ORBS slicer, and the Ochiai ranking strategy, which is one of the most effective ranking strategies in object-oriented programs \cite{Daming19,Jifeng14}.
Among the available repair tools, we choose TBar, a template-based APR tool, because this approach is more
effective at generating correct patches than other APR approaches.
However the approach suffers from efficiency issues making TBar highly
sensitive to the quality of the fault localisation algorithm and the size of test
suite~\cite{Liu20}.

\subsection{Benchmark Dataset and Repair Setups}

Our experiments use the Defects4J (v1.2.0) benchmark~\cite{JustJE14}, which is widely used for APR tasks~\cite{JiangXZGC18,LiuK0B19}. 
The benchmark contains a diversity of bug types~\cite{SobreiraDDMM18}.
For each of the examined subjects, we run TBar using the reduced test suites and suspicious lists and report the resultant RT and NPC values. 

\subsection{Experimental Results} \label{sec:empiricalResults}

\noindent
\headQ{RQ1: How effective is ORBS at removing irrelevant tests?}

\noindent \head{Setup}
We use the number of tests in a suite as a proxy for the cost of
executing the test suite.
This enables us to compute a \emph{reduction rate} as the ratio of the number of tests in the reduced suite to the number of tests in the original suite.

\noindent \head{Results}
Table~\ref{table:RQ1Results} shows the difference in the number of passing and failing test cases before and after reduction.
Several interesting patterns are evident in the Table~\ref{table:RQ1Results}.
First, as expected slicing does not affect the number of failing tests.
Thus all executions of the original program that exhibit bug $b$ have a corresponding execution of program slice that also exhibits $b$.

Second, the test suites that accompany Defects4J are designed to comprehensively test all functionalities of the corresponding software project.
This is indeed useful when generating a patch of a large software project,
because we typically do not know which parts of the project are affected by the
repair and thus need to validated it against the entire test suite to ensure
the program still satisfies its specification. 
However, our analysis shows that a large number of tests in test suites examine parts of the code that are irrelevant to the bug being repaired.
Thus, slicing can be an effective test suite minimisation technique to identify and remove irrelevant tests. 

\begin{table}
\centering
 \begin{tabular}{|c|c|c|c|c|} 
\hline
  & \multicolumn{2}{c}{\thead{Complete test suites}} & \multicolumn{2}{|c|}{\thead{Reduced test suite}} \\
\hline
\textbf{BugID} & \thead{Passing tests} & \thead{Failing tests} & \thead{Passing tests} & \thead{Failing tests} \\
\hline
Lang-10 & 2196 & 2 & 50 & 2  \\
\hline
Lang-39 & 1617 & 1 & 76 & 1  \\
\hline
Lang-44 & 1784 & 1 & 19 & 1  \\
\hline
Lang-45 & 1782 & 1 & 15 & 1  \\
\hline
Lang-51 & 1630 & 1 & 43 & 1  \\
\hline
Lang-58 & 1594 & 1 & 53 & 1  \\
\hline
Lang-59 & 1592 & 1 & 30 & 1  \\
\hline
\end{tabular}
\caption{The impact of ORBS on the size of test suites}
 \label{table:RQ1Results}
\end{table}

\begin{tcolorbox}
Answer to RQ1: 
ORBS is effective in identifying tests that are irrelevant to the bug being repaired. 
The integration of ORBS with APR reduces the size of test suites
without eliminating any relevant passing or failing test. In addition, the patch generated for each bug by the setup $R(P, T_R, SL)$ is identical to the one produced by $R(P, T, SL)$. 
This increases confidence in the soundness of test reduction provided by ORBS. 
\end{tcolorbox}

\headQ{RQ2: How effective is the ORBS at improving FL?}

\noindent \head{Setup}
APR tools are typically fed a ranked list of suspicious statements to be potentially mutated in turn.
Thus, fault localization's accuracy has a direct impact on the performance of an APR tool.
To evaluate the effectiveness of ORBS in improving the accuracy of
the fault localization, we run GZoltar under two setups: 
  (i) using original programs with complete test suite,
and 
  (ii) using original programs with reduced test suites. 

\noindent \head{Results}
Table~\ref{table:RQ2Results} summarizes the results.
First, for all seven versions the actual faulty statement is included in the slice.
This is crucial as it shows that slicing does not adversely affect the
capability of fault localization to localise the faulty statement.
Second, the table shows that ORBS improved the suspiciousness rank of the
faulty statement for six of the seven bugs. 
The reduction is roughly proportional to the rank of the faulty statement.
For example, for Lang-10 the suspiciousness rank improves 13 positions from 71 to 58, while for Lang-59 it improves only one position from 6 to 5. 



\begin{tcolorbox}
Answer to RQ2: combining GZoltar and  ORBS proved generally effective.
It reduced the size of suspicious lists and improved the suspiciousness rank of each relevant faulty statement. 
The advantage ORBS brings is more pronounced when the size of the suspicious list is large and rank of the faulty statement is low.
\end{tcolorbox}

\begin{table}[!htbp]
\centering
 \begin{tabular}{|c|c|c|c|c|}
\hline
  & \multicolumn{2}{c}{\thead{Original program}} & \multicolumn{2}{|c|}{\thead{Reduced program}}  \\
\hline
\textbf{BugID} & \thead{FL size} & \thead{Fault rank} & \thead{FL size} & \thead{Fault rank} \\
\hline
Lang-10 & 261 & 71 & 164 & 58 \\
\hline
Lang-39 & 51 & 27 & 42 & 24 \\
\hline
Lang-44 & 22 & 16 &  19& 14 \\
\hline
Lang-45 & 24 & 5 & 20 & 6 \\
\hline
Lang-51 & 19 & 10 &  16& 7 \\
\hline
Lang-58 & 54 &  20&  52& 19 \\
\hline
Lang-59 & 12 & 6 &  8&  5\\
\hline
\end{tabular}
\caption{The FL results using  original and reduced test suites} 
\label{table:RQ2Results}
\end{table}

\headQ{RQ3: Does improving FL  improve TBar's performance?}

\noindent \head{Setup}
From RQ2 ORBS improves the result of fault localization.
RQ3 asks if this improvement carries over to the performance of TBar.
To do so, we first run TBar using each original program, the complete test suite, and the original suspicious lists generated by the Ochiai ranking strategy. 
We then rerun TBar but substitute the optimised suspicious lists.
To isolate the impact of improved FL on the performance on APR, 
we hold the program $P$ and the test suite $T$ fixed while varying the list of
suspicious lines.

\noindent \head{Results}
Table~\ref{table:RQ3Results} compares the results of running TBar on the original buggy programs  using first the complete suspicious list
and then the reduced one.

In terms of NPC scores, combining ORBS and SBFL clearly helps.
For example, the suspiciousness rank of the faulty statement for Lang-10 improved from position
71 to position 58, which caused the NPC score to drop from 2027 to 1489. 
Overall the reduction in NPC score varies depending on how many positions a
faulty statement has been improved.

Recall that each suspicious statement in the generated suspicious list will be mutated using a number of fix templates, which varies depending on the syntactic structure of the statement.
A reduction in the NPC score should lead to a reduction in the total number of patch
candidates examined by TBar and thus a reduction on the time taken to find a valid patch.
From the table the impact of the NPC reduction on the repair time is significant, dropping the
 total repair time by more than half, from 7 hours to 3 hours and 20 minutes.

\begin{tcolorbox}
Answer to RQ3: 
the efficiency of TBar is influenced heavily by the accuracy of the FL approach.
Combining SBFL with ORBS we observed that the suspicious rank of each faulty statement improved and thus TBar's resultant NPC score was reduced. 
\end{tcolorbox}

\begin{table}[!htbp]
\centering
 \begin{tabular}{|c|c|c|c|}
\hline
\textbf{BugID} & \thead{Original NPC} & \thead{Reduced NPC}  & \thead{Reduction value}  \\
\hline
Lang-10 &  2027 &   1489 &   538\\
\hline
Lang-39 & 1350 &  1187 &  163   \\
\hline
Lang-44 &  438 & 357 & 81 \\
\hline
Lang-51 &   340 & 107  &  233\\
\hline
Lang-58 & 516 & 465 &   51 \\
\hline
Lang-59 &  4&  4&  0\\
\hline
\end{tabular}
\caption{The efficiency of TBar using NPC in original and reduced SLs}
 \label{table:RQ3Results}
\end{table}

\headQ{RQ4: Does reducing test suites  improve TBar's efficiency?}

\noindent \head{Setup}
Test suite minimization attempts to reduce testing cost be removing irrelevant tests.
When performing APR, test suite minimization aims to reduce the cost of repair
by reducing the cost of patch validation.
To address this, we run TBar under two different settings: 
(i) the original programs with complete tests,
and 
(ii) original programs with reduced tests while keeping the size of the suspicious list unchanged.

\head{Results}
Table~\ref{table:RQ5Results} shows the repair time cost for running TBar using the complete and the reduced test suites. 
As noted previously, NPC is not a good choice here because changing the size of
the test suite does not impact the number of patch candidates and thus the NPC
score stays the same.
Only the cost of patch validation where tests are run using generated patches is effected by changes in the test quite.

One clear trend in the data is that the performance of TBar is \emph{inversely proportional} to the rank of the faulty statement responsible for the target bug.
That is, the lower the suspiciousness rank of the faulty statement, the greater the impact of test suite reduction.
Note that as the faulty statement is ranked lower in the suspicious list, the number of examined patch candidates increases,
which in turn increases the number of patch compilations and test case executions.
Thus, by reducing the size of the test suite we obtain a great reduction on the overall processing time. 
For example, the test suite for Lang-10 dropped from 2198 to 52 tests, which reduced the 
repair time from 26340 to 5400 seconds. 
This is mainly due to the faulty statement of Lang-10 being ranked in the 71th
position in the original  suspicious list.

\begin{tcolorbox}
Answer to RQ4: Size of the test suite has a direct effect on TBar's efficiency.
This impact is proportional to the rank of the faulty statement responsible for the bug.
\end{tcolorbox}

\begin{table}[!htbp]
\centering
 \begin{tabular}{|c|c|c|c|} 
\hline
\textbf{BugID} & \thead{complete test suite}  & \thead{reduced test suite} &  \thead{RTC Reduction}    \\
\hline
Lang-10 & 26340  & 5400 &  21030  \\
\hline
Lang-39 &  11400 &  8880 & 2520 \\
\hline
Lang-44 &  360&  143& 217\\
\hline
Lang-45 & 4080 & 600  & 3480   \\
\hline
Lang-51 & 2760 &  480 & 2280 \\
\hline
Lang-58 & 1260 & 285 & 975    \\
\hline
Lang-59 & 34 & 25 & 9  \\
\hline
\end{tabular}
\caption{TBar's efficiency (in seconds) on complete and reduced suites}
 \label{table:RQ5Results}
\end{table}


\subsection{Reliable setups for the combination of slicing and APR}
This section provides a summary of empirically discovered reliable slicing-related repair setups. We consider a setup to be reliable if it produces a patch that is identical to the one produced by the setup $R (P, T, SL)$ or a patch valid for  program $P$.
Identifying reliable slicing-related repair setups is crucial for APR because it allows us to draw conclusions about the correctness of the patches created by the implemented slicing-related repair setup without needing to consider the standard repair setup $R (P, T, SL)$ with the complete test suite.

There are four setups to examine: $R (P_S, T_R, SL_R)$, $R (P, T_R, SL_R)$, $R (P, T_R, SL)$, and $R (P, T, SL_R)$.
However, during the analysis we observe cases where the setups $R (P, T_R, SL_R)$, $R (P, T_R, SL)$, and $R (P, T, SL_R)$ are able to produce valid patches for the analyzed bug while $R (P_S, T_R, SL_R)$ fails to find a patch (e.g., bug Math-5 in Defects4J). 
    This occurs, for instance, when the slicing operation deletes some expressions from the associated conditional branch, but the erroneous conditional statement itself is maintained in the generated slice. 
    This indicates that ORBS may delete statements that are necessary for generating patches.  

\begin{tcolorbox}
\head{Reliable repair setups} 
Empirical results demonstrate that the setups $R (P, T_R, SL_R)$, $R (P, T, SL_R)$, and $R (P, T_R, SL)$ give the same patch as the standard setup $R (P, T, SL)$ for all of the investigated defects.  On the other hand, the setup $R (P_S, T_R, SL_R)$ may fail to generate a valid patch for the original program. 
\end{tcolorbox}


\section{Related Work}
In this section, we discuss related work on template-based APR approaches and previous attempts on accelerating APR.
    
\head{Template-based APR}
Template-based APR are widely used in the APR literature~\cite{Durieux17,Hua18,JiangXZGC18,KoyuncuLBKKMT20,Bach16,Liu19,Liu18,LiuK0B19,LongAR17,Saha17,Wen18,Xin17},
which utilize predefined fix templates to fix specific bugs.
However, most of previous work on template-based APR has focused on maximizing the fix-rate 
by incorporating a large number of useful fix templates.
Despite of these great achievements, modern template-based APR tools
still suffer from the efficiency issue as demonstrated in~\cite{Liu20}  and this work. 
Combining slicing with APR would help not only in improving
the efficiency of the repair process by reaching faulty statements earlier, but also increasing their capabilities in producing correct patches.

\head{Previous attempts on accelerating APR}
Several techniques have been developed to increase the performance of APR,
including regression test selection~\cite{YooH12,Mehne18}, patch filtering~\cite{Liang21,Yang21},
and on-the-fly patch generation~\cite{Hua18} and validation~\cite{Chen21,Benton22}. 
The goal of these techniques is to  reduce the patch compilation and test case execution costs,
which are the dominant contributors for APR runtime. 
While these techniques have the potential to reduce the repair cost of programs, they affect a particular step of the APR process.
Combining slicing with APR has the potential to improve all aspects of APR, including the fault localization step, patch generation step, and patch validation step. 

\section{Conclusion and Future Work}

Repairing large-scale programs is a time-consuming process. 
The most challenging parts of the APR process are finding faulty statements and verifying patches.
Instead of examining the entire program without any useful hints, this work proposed to focus on a relatively small portion of the program containing suspicious statements that directly cause program bugs. 
This can be achieved by applying program slicing  ahead of the repair process.
The experiments show  that the effect of slicing is mostly positive and can significantly improve the performance of APR tools. 
However, we need exhaustive
experiments involving programs with various program sizes
and bug types to confirm the general effectiveness of the
approach. We aim to report on this in future work.

    
\section*{Acknowledgements}    

The author would like to acknowledge Dave Binkley for providing the slices for the Lang project and Linas Vidziunas for the help with the experimental setup. The author completed this work while working at the Simula Research Laboratory.

\bibliographystyle{IEEEtran}
\bibliography{IEEEabrv,references}

\end{document}